\documentclass[11pt]{article}

\let\ssection=\section
\renewcommand{\section}{\setcounter{equation}{0}\ssection}

\newcommand{\half}{{\scriptstyle{\frac{1}{2}}}}

\newcommand{\vx}{{\vec x}}

\def\p{{\partial}}

\def\vp{{\vec p}}

\def\vA{{\vec A}}
\def\vnabla{{\vec\nabla}}
\newcommand{\FS}{F\cdot S}

\begin{document}

\setlength{\baselineskip}{16pt}

\title{Non-commuting coordinates,  exotic particles,
\& anomalous anyons 
in the Hall effect\footnote{Talk given by P. Horv\'athy at
the International Workshop 
{\it Nonlinear Physics: Theory and Experiment.}\ {\rm III}. 
July'04, Gallipoli, (Lecce, Italy). To be published in
{\it Theor. Math. Phys}.
}}

\author{ 
 C.~Duval 
 \\ 
 Centre de Physique Th\'eorique, CNRS\\ 
 Luminy, Case 907\\ 
 F-13288 MARSEILLE Cedex 9 (France)\\ 
 (UMR 6207 du CNRS associ\'ee aux 
 Universit\'es d'Aix-Marseille I et II \\ 
 et Universit\'e du Sud Toulon-Var; laboratoire 
 affili\'e \`a la FRUMAM-FR2291) 
 \\[8pt] 
 P.~A.~Horv\'athy 
 \\ 
 Laboratoire de Math\'ematiques et de Physique Th\'eorique\\ 
 Universit\'e de Tours\\ 
 Parc de Grandmont\\ 
 F-37200 TOURS (France) 
 \\ 
 } 
\date{\today}

\maketitle

\begin{abstract}
      Our previous ``exotic'' particle, together with the more
      recent anomalous anyon model (which has arbitrary gyromagnetic 
      factor $g$) are reviewed. The non-relativistic limit of the anyon
      generalizes the exotic particle which has $g=0$ to any $g$.
      When put into planar electric and magnetic fields,
      the Hall effect becomes mandatory for all $g\neq2$, 
      when the field takes some critical value.
\end{abstract}

\texttt{hep-th/0407010}

\section{Introduction}

Classically, charged particles in the Landau problem move along helical 
paths around the guiding center, which follows in turn the Hall law. 
Quantum mechanically, restriction to guiding center motion amounts to
projecting to the Lowest Landau Level (LLL).
The guiding center coordinates are 
non-commuting \cite{LaLi, GirvinJach, Ezawa},
\begin{equation}
    \big[Q_{1},Q_{2}]=\frac{i}{eB}.
    \label{NCcoord}
\end{equation}
  
Following Laughlin, the Fractional Quantum Hall 
Effect (FQHE), observed in thin films of semiconducting 
heterostructures, should be explained within the LLL \cite{QHE}.
In a more refined picture, the charges become
 quasiparticles, namely charged planar vortices \cite{QHE,Ezawa}.
 The dynamics of the latter can in turn be
described  by a simple Hamiltonian model \cite{DJT,vortmot}.
The coordinates are once again non-commuting. This  
model can actually be obtained by reduction
of an ``exotic particle'' \cite{DH}, associated with
the two-fold central extension of the planar Galilei group 
\cite{exotic},
which has non-commuting coordinates from the outset \cite{DH}.
For a critical value of the magnetic field, (\ref{criticalcase}), 
the only consistent motions follow the Hall law \cite{DH}.
Quantization allowed us
to derive the wave functions Laughlin starts with,
and the commutation relation (\ref{NCcoord}) is recovered \cite{DH}.

Although we only have circumstantial evidence but
no convincing proof yet, 
we believe that our ``exotic particles'' should describe
the effective motion of Laughlin's quasiparticles.

Mathematically, the ``exotic'' model arises due to the particular
properties of the plane, namely the commutativity of 
rotations \cite{exotic}. But what is its physical origin~? 
Jackiw and Nair \cite{JaNa} have obtained the free ``exotic''
model  as a non-relativistic limit of anyons\footnote{By an
``anyon'' we simply mean a particle in the plane 
carrying (arbitrary) spin.}, whereas the second extension parameter
of the Galilei group
arises as a ``non-relativistic shadow'' of relativistic spin \cite{anyons}.

Coupling anyons to an electromagnetic field has been 
considered \cite{CNP,anyong}, and it has been 
claimed \cite{JaNa} that
our coupled exotic system would be in fact the ``Jackiw-Nair'' (JN)
limit of the one studied in Ref. \cite{CNP}.  
That this cannot true, is seen by comparing the
 gyromagnetic ratios. The model in \cite{CNP}  has $g=2$; 
in fact, some high-energy theoreticians \cite{CNP,anyong} have argued 
that the gyromagnetic ratio of anyons is necessarily $g=2$. 
This contradicts experimental
evidence, though~:  in the FQHE $g\sim 0$  \cite{Ezawa,g0}.

In a recent paper \cite{AnAn} a generalized anyon model was presented,
which accomodates any value of the gyromagnetic ratio. It was 
found that, when $g\neq2$, all motions follow the Hall law
{\it provided} the field and the spin satisfy a certain relation
which generalizes (\ref{criticalcase}).
Non-relativistic counterparts can be
derived by taking the JN limit.
The ``exotic'' model of Ref. \cite{DH} is recovered, in particular,
when $g=0$.

\section{Exotic particles}

The ``exotic'' model of Ref. \cite{DH}, associated with the
two-fold centrally extended Galilei group \cite{exotic}
has been constructed \cite{DH,Grigore} by Souriau's group
theoretical method \cite{SSD}.
Expressed in a more conventional language, the four dimensional
phase space is endowed with the ``exotic'' symplectic structure and
 free Hamiltonian,
\begin{equation}
    \Omega_{0}=d\vp\wedge d\vx+\frac{\theta}{2}\epsilon_{ij}dp_{i}\wedge dp_{j},
    \qquad
    H_{0}=\frac{\vp{}^2}{2m},
\label{exoticsymp}
\end{equation} 
respectively. Note that the second, ``exotic'' term in
the symplectic form only exists in the plane.
By construction, the system (\ref{exoticsymp}) realizes the
two-fold centrally extended Galilean symmetry \cite{exotic,DH}.
Energy and momentum are conventional;
the exotic structure only appears in
the angular momentum and Galilean boosts,
\begin{equation}
    \begin{array}{ll}
    &j=\vx\times\vp+\half\theta\,\vp^2,
    \\[6pt]
    &K_{i}=mx_{i}-p_{i}t+m\theta\,\epsilon_{ij}p_{j}.
    \end{array}
\label{consquant}
\end{equation}
The commutation relations are the usual ones except for boosts, 
which satisfy rather
\begin{equation}
    \big\{K_{1}, K_{2}\big\}=-m^2\theta,
\end{equation}
which is the hallmark of ``exotic'' Galilean symmetry \cite{exotic}.
Minimal coupling is achieved by adding the electromagnetic two-form 
$eF$  \cite{SSD}.
This yields the fundamental Poisson brackets 
\begin{equation}
\begin{array}{lll}
\{x_{1},x_{2}\}=
\displaystyle\frac{m}{m^*}\,\theta,
	\\[3mm]
	\{x_{i},p_{j}\}=\displaystyle\frac{m}{m^*}\,\delta_{ij},
	\\[3mm]
	\{p_{1},p_{2}\}=\displaystyle\frac{m}{m^*}\,eB,
\end{array}
\label{exocommrel}
\end{equation}
where 
\begin{equation}
    m^*=m(1-\theta eB).
    \label{effmass}
\end{equation}

Our commutation relations (\ref{exocommrel}) are similar to
but still different from those posited later by Nair and 
Polychronakos \cite{NaPo},
\begin{equation}
\begin{array}{lll}
\{x_{1},x_{2}\}=\theta,
	\\[3mm]
	\{x_{i},p_{j}\}=\delta_{ij},
	\\[3mm]
	\{p_{1},p_{2}\}=eB.
\end{array}
\label{NaPocommrel}
\end{equation}
The two  systems are in fact only equivalent for a free particle, $F=0$.

Superficially, the relations (\ref{NaPocommrel}) look more natural. 
They are, however, only consistent
for a homogeneous magnetic field, $\vnabla B=0$; otherwise, the
Jacobi identity is violated \cite{DH}. E.g., 
\begin{equation}
\big\{x_{i},\{p_{1},p_{2}\}\big\}_{\rm cyclic}
 =\theta\varepsilon_{ij}\p_{j}B.
\end{equation}

The  strange-looking ``exotic'' relations (\ref{exocommrel}) only
require in turn the natural condition $dF=0$, and have therefore
a broader validity.

The difference of the models comes from the coupling:
while we use Souriau's prescription (adding two-forms), 
the authors of  \cite{NaPo} posit the Poisson brackets (\ref{NaPocommrel}).
It is intriguing to observe that the models can nevertheless
can be carried into each other, namely by redefining the time
as
\begin{equation}
t \to \frac{m}{m^*}\,t,
\end{equation}
This is allowed as long as $B$ is constant such that $m^*\neq0$.
For $m^*=0$ it breaks down, however, yielding inequivalent
behaviour.

In what follows, we only consider the more satisfactory ``exotic'' model.

In a Lagrangian framework, our minimally coupled system is 
described by the first-order expression \cite{DH}
\begin{equation}
    L=(\vp-e\vA)\cdot\dot{\vx}-\frac{\vp^2}{2m}
    -eV+\frac{\theta}{2}\,\vp\times\dot{\vp}.
\label{mincouplag}
\end{equation}
 In the associated Euler-Lagrange  (or, equivalently, 
 Hamilton) equations,
\begin{equation}
    \begin{array}{ll}
    &m^*\dot{x}_{i}=p_{i}-em\theta\epsilon_{ij}E_{j},
    \\[4pt]
    &\dot{p}_{i}=eE_{i}+eB\epsilon_{ij}\dot{x}_{j},
    \end{array}
\label{mincoupleqmot}
\end{equation}
the extension parameters combine with the magnetic field
into an {\it effective mass}, namely $m^*$ in (\ref{effmass}).
For non-vanishing effective mass, $m^*\neq0$,
the motions are roughly similar to that of an ordinary particle
in a planar electromagnetic field.
When the effective mass vanishes, $m^*=0$, i. e. for
\begin{equation}
    B=\frac{1}{e\theta},
    \label{criticalcase}
\end{equation}    
however, the system becomes singular, and the
consistency of the equations of motion (\ref{mincoupleqmot}) can 
only be maintained if {\it the guiding center coordinates},
\begin{equation} 
    Q_{i}=x_{i}-\frac{mE_{i}}{eB^2},
    \label{guidingcenter} 
\end{equation}
{\it follow  the  Hall law},
\begin{equation}
    \dot{Q}_{i}=\varepsilon_{ij}\frac{E_{j}}{B},
\label{Hallaw}
\end{equation} 
where $E_{i}=-\partial_{i}V$.
The proof goes as follows. Put 
$$
\widetilde{Q}_{i}=x_{i}+\epsilon_{ij}\frac{p_{j}}{eB}
\qquad\Longrightarrow\qquad
\dot{\widetilde{Q}}_{i}=
\dot{x}_{i}+\epsilon_{ij}\frac{\dot{p}_{j}}{eB}.
$$
Inserting here the velocity expressed from the Lorentz force law,
$
\dot{x}_{i}=\epsilon_{ij}{E_{j}}/{B}-\epsilon_{ij}{\dot{p}_{j}}/{eB},
$
the $\dot{p}_{j}$ terms cancel, leaving us with (\ref{Hallaw})
for $\widetilde{Q}$
\footnote{This also holds in the commutative case.}. 
What is specific for us here is that,
for $m^*=0$, the first equation in (\ref{mincoupleqmot}) requires
\begin{equation}
    p_{i}=m\,\epsilon_{ij}\frac{E_{j}}{B},
    \label{momconstraint}
\end{equation}
Momentum is, therefore, no longer dynamical, since it is determined 
by the position. 
The only dynamical degrees of freedom are the
$\widetilde{Q}_{i}$ [in fact
 $\widetilde{Q}_{i}=Q_{i}$ by (\ref{momconstraint})],
whose (Hall) motion determines that of the physical coordinate
$x_{i}$. This realizes Laughlin's idea, 
who argued that the FQHE should amount to ``condensation
into a collective ground state'' \cite{QHE}.
\goodbreak
 
For a constant electric field in particular, 
the additional term in (\ref{guidingcenter}) is just a constant, and
the particle itself follows the Hall law.

Condition (\ref{criticalcase})
is consistent with (\ref{NCcoord}) in that it fixes the
value of the magnetic field 
 as a function of the non-commutative parameter $\theta$,
considered as a physical parameter of the particle on the same footing
as mass. For this value projection to the LLL is mandatory.

The intuitive picture is the following.
For $m^*\neq0$, the particle moves along helical trajectories 
around the guiding center.
These latters can actually be genuine motions for specific
initial conditions, namely when the initial velocity is such that the
Lorentz force is precisely cancelled by the electric force.
Then the motion is along equipotentials. When $m^*\to0$, however,
we are left with the guiding center motion alone, since 
the initial conditions of would-be helical motions become
forbidden. Their inconsistency results in instantaneous propagation
\cite{DH} that makes the quantum propagator finite for infinitesimal
time \cite{GirvinJach}.

\section{Relativistic anyons}\label{CNPmod}

The ``exotic'' extension plays hence little r\^ole for a free
particle, and only becomes important when interactions
are considered. This reminds one of what happens for a
spinning particle. The similarity is not accidental:
as it has been pointed out by Nair and
Jackiw \cite{JaNa}, the free ``exotic'' model can indeed
be obtained as a tricky non-relativistic limit of a
free, relativistic anyon. In detail, the latter is descibed
by the symplectic form and Hamiltonian \cite{anyons} \footnote{
Greek indices $\alpha, \beta$, etc.
denote $2+1$ dimensional coordinates and 
$i,j$, \dots are spatial coordinates,
respectively. Our metric is diag$(c^2,-1,-1)$.},
\begin{equation}
\Omega_{0}=dp_\alpha\wedge dx^\alpha+\frac{s}{2}
\epsilon^{\alpha\beta\gamma}\,\frac{p_\alpha{}dp_\beta\wedge{}dp_\gamma}
{(p^2)^{3/2}},
\qquad
H_{0}=\frac{1}{2m}\left(p^2-m^2c^2\right).
\label{JNsymp}
\end{equation} 
Then, expanding as $p_{0}\simeq  mc^2+\vp^2/2m$, putting
\begin{equation}
    s/c^2=m^2\theta
\end{equation}
 and only keeping leading terms in $c^{-2}$ yields
precisely the free ``exotic'' structure.

Let us first review the theory of 
Chou et al. \cite{CNP}. These authors {\it posit}, without further
justification,
that anyons in {\it weak and constant} electromagnetic field
 move following the usual (planar) Lorentz equations
\begin{eqnarray}
	m\displaystyle\frac{dx^\alpha}{d\lambda}=p^\alpha,
	\qquad	
	\displaystyle\frac{dp^\alpha}{d\lambda}=\frac{e}{m}
	F^{\alpha\beta}p_{\beta}
	\label{CNPeq}
\end{eqnarray}
where $F_{\alpha\beta}$ is the electromagnetic field, $e$ the charge 
and $m$ the mass; $\lambda$ denotes proper time.
 These equations are associated with the symplectic structure and 
Hamiltonian on $6$-dimensional phase space
\begin{eqnarray}
    \Omega&=
    &dp_{\alpha}\wedge dx^{\alpha}
    +\displaystyle\frac{s}{2}
\epsilon^{\alpha\beta\gamma}\,
\displaystyle\frac{p_\alpha{}dp_\beta\wedge{}dp_\gamma}
{(p^2)^{3/2}}
    +
    \half eF_{\alpha\beta}dx^\alpha\wedge dx^\beta,\label{CNPsymp}
    \\[6pt] 
    H&=&\displaystyle\frac{1}{2m}\left(p^2-m^2c^2
    +\frac{es}{\sqrt{p^2}}
    \varepsilon_{\alpha\beta\gamma}F^{\alpha\beta}p^\gamma\right),
    \label{CNPHam}
\end{eqnarray}
respectively. Curiously,  
for each value of $s$ one gets the same equations of motion, namely 
(\ref{CNPeq}).
Note that while the symplectic form $\Omega=\Omega_{0}+eF$ is obtained 
by the familiar  minimal coupling prescription \cite{SSD},
 the Hamiltonian is  modified by the addition of a  
 non-minimal term, $H=H_{0}+H_{s}$, where $H_{s}$ is {\it chosen} so 
as to cancel the effect of the spin term in the symplectic
form. Thus, its r\^ole is to enforce the posited relation 
(\ref{CNPeq}) between $p_{\alpha}$ an $\dot{x}_{\alpha}$.

In the plane, spin is basically reduced to a real constant.
If we require, as usually, that the spin tensor
$S_{\alpha\beta}$ satisfy the condition 
\begin{equation}
    S_{\alpha\beta}p^\beta=0,
    \label{spinconstr}
\end{equation}
it follows that spin and momentum are proportional,
\begin{equation} 
    S_{\alpha\beta}=\frac{s}{\sqrt{p^2}}\,
    \epsilon_{\alpha\beta\gamma}p^{\gamma}.
    \label{spintensor}
\end{equation} 
Let us  observe for further reference that, introducing the shorthand
$
\FS=-F_{\alpha\beta}S^{\alpha\beta}
$, 
the additional term picked by Chou et al. can be written in the form
$
H_{s}=-({e}/{2m})\FS.
$ 
Their Hamiltonian in (\ref{CNPHam}) is therefore 
\begin{equation}
     H=\frac{1}{2m}\big(p^2-M^2c^2\big)
     \quad\hbox{where}\quad
M^2=m^2+\frac{e}{c^2}\,\FS.
\label{oldm2}
\end{equation} 
Here $M$ can be viewed as a {\it field-dependent mass},
which also includes the magnetization energy \cite{JMS}.
Then solving Hamilton's equations associated with $\Omega$ and $H$
are in fact the same as finding the kernel of $\Omega$
restricted to the $5$ dimensional surface defined by $H=0$ i.e.
$
p^2=M^2c^2.
$ 

\section{A generalized model}

Recently, a more general model has been proposed \cite{AnAn}.
The clue is to generalize the ``field-dependent mass''
$M$ in (\ref{oldm2}) by including a $g/2$ factor~: 
we still choose $H=p^2-M^2c^2$, but with
\begin{equation}
M^2=m^2+\frac{g}{2}\cdot\frac{e}{c^2}\,\FS.
\label{newm2}
\end{equation}
This yields \cite{AnAn} the generalized equations
\begin{eqnarray}
D\,  \displaystyle\frac{dx^\alpha}{d\tau}
&=&
G\, \frac{p^{\alpha}}{M}
+(g-2)\frac{es}{4M^2}
\epsilon^{\alpha\beta\gamma}
F_{\beta\gamma},\label{vitesse}
\\[8pt]
\displaystyle\frac{dp^{\alpha}}{d\tau}
&=&eF^\alpha_{\ \beta}\displaystyle\frac{dx^{\beta}}{d\tau},
\label{Lorentzforce}
\end{eqnarray}
where
\begin{equation}
     D=1+\frac{e\FS}{2M^2c^2},
     \qquad
     G=1+\frac{g}{2}\cdot\frac{e\FS}{2M^2c^2}.
     \label{factors}
\end{equation}
     
A remarkable property is that, for $g\ne2$, velocity and momentum are
{\it not} parallel. Such a possibility has been 
advocated a long time ago \cite{Papa}.
For $g\ne2$, our generalized model has a remarkable behaviour:
when either of the coefficients $D$ or $G$ vanish, 
the {\it only allowed motions} are governed by the
Hall law \cite{AnAn}. When $g=2$ instead, $D=G$ drops 
out and the Hall effect disappears. 

We believe, therefore, that
the Hall Effect is a physical manifestation of anomalous anyons.

Our new coupling defined through
the field-dependent mass in (\ref{newm2}) is in general non-minimal:
 minimal coupling corresponds to 
{\it anomalous gyromagnetic factor} $g=0$ and {\it not} to
$g=2$.

Further insight can be gained calculating the JN
limit of the relativistic anyon. Easy calculation \cite{AnAn}
shows indeed that (\ref{vitesse}-\ref{Lorentzforce}) go over into
\begin{eqnarray}
     \Big(1-(g+1)\theta eB\Big)\dot{x}^i&=&
     \Big(1-\frac{3g}{2}\theta eB\Big)\frac{p^i}{M_{g}}
     -\left(1-\frac{g}{2}\right)e\theta\epsilon^{ij}E_{j}\label{NRvitesse}
     \\[8pt]
     \dot{p}^i&=&eE^{i}+eB\epsilon^{ij}\dot{x}_{j}
     \label{NRLorentzforce}
\end{eqnarray}
where the dot means derivation w.r.t. nonrelativistic time
and $M_{g}=m\sqrt{1-g\theta eB}$ is the non-relativistic limit of
the field-dependent mass $M$.
Then, it is now easy to see that when either
\begin{equation}
     B'=\frac{1}{1+g}\cdot\frac{1}{e\theta}
     \qquad\hbox{or}\qquad
     B''=\frac{2}{3g}\cdot\frac{1}{e\theta},
     \label{NRcritB}
\end{equation}
the particle must obey the Hall law. The first of these
critical values generalizes (\ref{criticalcase}) to any $g$,
whereas the second is of a novel type.

For $g=0$, we recover in particular the ``exotic'' model
of \cite{DH}. 
This latter can {\it not} be, therefore, the NR limit of the
Chou et al. system, which has $g=2$.

For $g=2$ our equations (\ref{vitesse}-\ref{Lorentzforce}) 
can be mapped into those proposed by Chou, et al. in  \cite{CNP}
by reparametrizing as
\begin{equation}
\tau\to \frac{m}{M}\,\tau
\end{equation}
where
$\tau$ is proper  time and $M$ is our (relativistic)
field-dependent mass (\ref{newm2}).
 It is worth mentioning out that our mass formula (\ref{newm2}) is 
{\it not} mandatory: {\it any} function $M(\phi)$, $\phi=\FS$ 
of the spin-field coupling
would provide us with a consistent theory \cite{JMS,AnAn}.
Other choices have also been considered \cite{JMS,Kunzle}. 
K\"unzle \cite{Kunzle}, for example, chose
\begin{equation}
\widetilde{M}=m+\frac{ge}{4mc^2}\,\FS.
\label{Km2}
\end{equation}

A general mass function $M(\FS)$ would yield again 
(\ref{vitesse}-\ref{Lorentzforce}), with gyromagnetic factor  
\begin{equation} 
\frac{g}{2}=\frac{c^2}{e}\,\frac{d(M^2)}{d\phi}.
\end{equation} 
Again, when the system becomes either
singular, $D=0$, or when the momentum uncouples, $G=0$,
all motions obey the Hall law, provided $g\neq2$.

The choices (\ref{newm2}) and (\ref{Km2})
lead to the same equations in the weak-field limit. In fact
$M=\sqrt{m^2+eg\FS/2c^2}\simeq \widetilde{M}$ if $eg\FS/m^2c^2<<1$.

Although in this paper we focused our attention to the behaviour
of a charged particle in an electromagnetic field, its is worth
mentioning that a similar study would also be possible for 
gravitational interactions \cite{JMS, Kunzle}. 
In a very strong gravitational field,
in particular, one could reveal analogous degenerate motions
\cite{Kunzle}. Similar equations have been obtained 
in a Lagrangian framework \cite{Scripta}.
\goodbreak

\noindent{\bf Acknowledgments}.
 Discussions with L. Martina, M. Plyushchay and
P. Stichel are acknowledged.
PH would like to thank the organizers of
International Workshop on {\it Nonlinear Physics: Theory and Experiment. 
}{\rm III}. {\it Gallipoli, (Lecce, Italy)}
for their hospitality during the Workshop.

\goodbreak

\end{document}